\documentclass[english,superscriptaddress,preprint,prb]{revtex4}
\usepackage[T1]{fontenc}
\usepackage[latin9]{inputenc}
\usepackage{color}
\usepackage{babel}
\usepackage{units}
\usepackage{mathtools}
\usepackage{amsmath}
\usepackage{graphicx}
\usepackage{setspace}

\usepackage[unicode=true,bookmarks=true,bookmarksnumbered=true,bookmarksopen=false,colorlinks=true]{hyperref}
\hypersetup{pdftitle={A unified description of magnetic field, angular, frequency and k-dependent collective magnetic excitations}, pdfauthor={Benjamin W. Zingsem}}
\usepackage{breakurl}

\begin{document}

\title{A unified description of 
collective magnetic excitations}

\author{Benjamin W. Zingsem}
\affiliation{Faculty of Physics and Center for Nanointegration (CENIDE),\\
	University Duisburg-Essen, 47057 Duisburg, Germany}
\affiliation{Ernst Ruska-Centre for Microscopy and Spectroscopy with Electrons and Peter Gr\"unberg Institute,\\
	Forschungszentrum J\"ulich GmbH, 52425 J\"ulich, Germany}

\author{Michael Winklhofer}
\affiliation{Faculty of Physics and Center for Nanointegration (CENIDE),\\
	University Duisburg-Essen, 47057 Duisburg, Germany}
\affiliation{IBU/School of Mathematics and Science, University of Oldenburg,\\
	Carl-von-Ossietzky-Strasse 9-11, D-26129 Oldenburg, Germany}
	
\author{Ralf Meckenstock}
\affiliation{Faculty of Physics and Center for Nanointegration (CENIDE),\\
	University Duisburg-Essen, 47057 Duisburg, Germany}
	
\author{Michael Farle}
\affiliation{Faculty of Physics and Center for Nanointegration (CENIDE),\\
	University Duisburg-Essen, 47057 Duisburg, Germany}
\affiliation{Center for Functionalized Magnetic Materials, Immanuel Kant Baltic Federal University,\\
	236041 Kaliningrad, Russian Federation}

\begin{abstract}
In this work, we define a set of analytic tools to describe the dynamic response of the magnetization to small
perturbations, which can be used on its own or in combination with micromagnetic 
simulations and does not require saturation. We present a general analytic description of the ferromagnetic high
frequency susceptibility tensor to describe angular
as well as frequency dependent ferromagnetic resonance spectra and account for asymmetries
in the line shape. Furthermore, we expand this model to reciprocal
space and show how it describes the magnon dispersion. Finally
we suggest a trajectory dependent solving tool to describe the equilibrium
states of the magnetization. 
\end{abstract}

\maketitle

\section{Introduction}
Many solutions of the Ferromagnetic high frequency susceptibility
(Polder-) tensor \cite{doi:10.1080/14786444908561215} 
have been formulated. Mostly these solutions represent simplified versions to
suit particular problems, such as a certain energy landscape, a certain
kind of coupling or a specific symmetry. In this work we formulate
a generalized linearization of the Landau-Lifshitz-Gilbert \cite{Landau1935,Gilbert2004} equations
(LLG), which does not require symmetry assumptions and is applicable regardless
of the coupling as well as the types of damping present in the system. It allows one to start with a general formulation of the free energy density of ferromagnets, including all magnetic inderactions which might be present in the magnetic material, e.g. exchange, dipole-dipole, Dzyaloshinskii-Moriya interaction, anisotropies, etc. and can be expanded to antiferromagnetis and multilayer artificial antiferromagnets in the usual way \cite{Kittel1951AntifRes,Keffer1952AntifRes,Vonsovskiui1966}.
The conventional approaches mostly involve solving large systems of
equations, linearizing at different points in the calculation
in order to formulate the high frequency susceptibility \cite{PhysRev.105.74,Suhl1955,doi:10.1080/14786444908561215}. This is avoided
here, by applying a straight forward linearization through series
expansion of the LLG. Furthermore this algorithm is formulated to
cover the entire magnon dispersion, including ferromagnetic resonance
modes as well as traveling waves with non-zero wave-vectors.

In the second part we present a model that can be used to calculate
the equilibrium orientations of the magnetization, using an algorithm
that closely resembles the actual measurement procedures used in ferromagnetic
resonance measurements. Following a gradient of the energy landscape imposed on the magnetization, this model can be used
to describe meta-stable and stable equilibrium states of the magnetization even for fields that are not applied along symmetry directions.

Neglecting thermal fluctuations, a combination of those models can
be used to make accurate predictions about the magnetodynamic properties
of ferromagnetic systems.

\section{Analytic Model}

\subsection{The ferromagnetic high frequency susceptibility tensor}

In the derivation presented here we assume a system that is described
by one macro-spin \cite{doi:10.1080/14786444908561215,Vonsovskiui1966,PhysRev.105.74,Suhl1955,smit1955philips,0034-4885-61-7-001} $\vec{M}$ 
which is subjected to one effective magnetic
field $\vec{B}$ yielding one high frequency susceptibility tensor
$\underline{\underline{\chi}}_{\mathrm{hf}}$. This model therefore
as derived here is designed to describe a fully saturated sample.
It is not limited to a single magnetization though and can be applied
to a set of macro-spins where the local effective magnetic field is known at each site. In that case the total high frequency susceptibility would be given as $\chi=\underset{n}{\sum}\chi_{n}$ where $\chi_{n}$ is
the high frequency susceptibility of the $n^{\mathrm{th}}$ macro-spin
$\vec{M}_{n}$ due to the field $\vec{B}_{n}$ it is exposed to. This
can be used for non saturated systems and samples with inhomogeneous
magnetization or magnetic nanoparticle configurations.

\begin{figure}
\noindent \begin{centering}
\includegraphics[width=0.5\columnwidth]{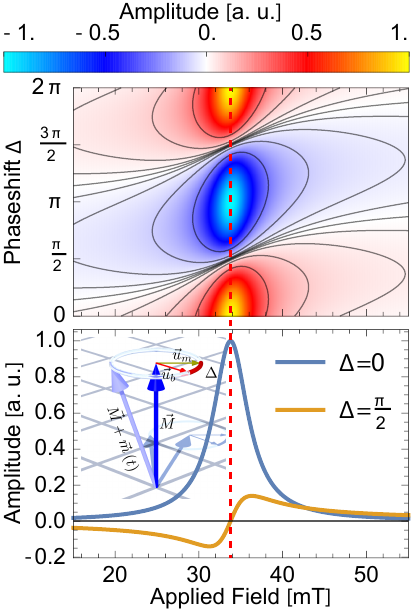}
\par\end{centering}
\caption{Top: Amplitude of the susceptibility as a function of the applied
magnetic field and the phase-shift $\Delta$. At the angles $\nicefrac{\pi}{2}$
and $\nicefrac{3\pi}{2}$ the line-shape is fully anti-symmetric similar
to the derivative of a Lorenz line-shape. In the vicinity of those
angles the signal is asymmetric. At the angles $0$ and $2\pi$ the
signal is symmetric with positive amplitude and at $\pi$ the signal
is symmetric with negative amplitude. Bottom: Selected
lines at specific angles $0$ and $\frac{\pi}{2}$. The inset illustrates the phase
shift $\Delta$ induced by choosing a nonparallel tuple of excitation
measurement projection vectors $\vec{u}_{b}$ and $\vec{u}_{m}$ shown
in the precession cone of the time dependent magnetization. For simplicity
the the precession is indicated as a circular motion normalized to
the length of the unit vectors perpendicular to $\vec{M}$. In general
it would be approximated to be elliptical and the opening of the cone
is much smaller compared to the Magnetization vector.\label{fig:Amplitude-of-the}}
\end{figure}

In order to derive the full tensor we start from the Landau-Lifshitz-Gilbert
Equation \ref{eq:LLG} using the Polder-Ansatz \cite{doi:10.1080/14786444908561215}
as shown in eq. \ref{eq:Ansatz-Polder} 
\begin{equation}
\vec{L}\coloneqq-\gamma\vec{M}\times\vec{B}-\frac{\alpha}{M}\vec{M}\times\dot{\vec{M}}-\dot{\vec{M}}=0\label{eq:LLG}
\end{equation}
\begin{equation}
\begin{array}{ccc}
\vec{M}\left(t\right) & \coloneqq & \vec{M}\left(M,\theta_{M},\varphi_{M}\right)+\vec{m}\exp\left(\imath\omega t\right)\\
\vec{B}\left(t\right) & \coloneqq & \vec{B}\left(B,\theta_{B},\varphi_{B}\right)+\vec{b}\exp\left(\imath\omega t\right)
\end{array}\label{eq:Ansatz-Polder}
\end{equation}
Considering the dynamic excitation and response quantities $\vec{m}$
and $\vec{b}$ to be sufficiently small, the ferromagnetic high-frequency
susceptibility $\underline{\underline{\chi}}_{\mathrm{hf}}$ can be
expressed as a linear tensor 
\begin{equation}
\vec{m}=\underline{\underline{\chi}}_{\mathrm{hf}}\cdot\vec{b}\label{eq:Hf-Tensor-Def}
\end{equation}
where linear means, that $\underline{\underline{\chi}}_{\mathrm{hf}}$
does not depend on $\vec{m}$ and $\vec{b}$. This is usually the
case for microwave fields $\vec{\left\Vert b\right\Vert }<\unit[1]{mT}$.
To obtain the magnetic flux that the magnetization is exposed to,
we consider the magnetic contribution to the free energy per unit volume
$F\left(\vec{B}_{\mathrm{appl}},\vec{M}\right)$ where $\vec{B}_{\mathrm{appl}}$
corresponds to the applied magnetic field and $\vec{M}$ is the magnetization
vector as discussed in the literature (See for example \cite{0034-4885-61-7-001}).
The Helmholtz free energy density $F$ usually contains an anisotropic
contribution due to the crystal lattice, particularly spin orbit interaction,
as well as several other contributions that arise from surfaces/interfaces,
the shape of the sample and the Zeeman-Energy. In this generalized
approach the nature of these magnetic energies almost does not matter. The only
necessary requirement is that the first and second derivatives used
in eq. \ref{eq:B-Feld_full_dynamic} exist. The total magnetic flux
is then given as 
\begin{equation}
\begin{array}{c}
\vec{B}(t)=\nabla_{\vec{M}}F\left(\vec{B}_{\mathrm{appl}},\vec{M}\right)\\
+\underline{\underline{J}}_{\vec{M}}\left(\nabla_{\vec{M}}F\left(\vec{B}_{\mathrm{appl}},\vec{M}\right)\right)\cdot\vec{m}\exp\left(\imath\omega t\right)\\
+\vec{b}\exp\left(\imath\omega t\right)
\end{array}\label{eq:B-Feld_full_dynamic}
\end{equation}
where $\nabla_{\vec{M}}F\left(\vec{B}_{\mathrm{appl}},\vec{M}\right)$
is the anisotropy-field and $\underline{\underline{J}}_{\vec{M}}\left(\nabla_{\vec{M}}F\left(\vec{B}_{\mathrm{appl}},\vec{M}\right)\right)$
the response function that accounts for a field caused by a precessing
$\vec{m}$, where $\nabla_{\vec{M}}$ is the gradient in $\vec{M}$
and $\underline{\underline{J}}_{\vec{M}}$ the Jacobian matrix in
$\vec{M}$. Using this we can now go back to eq. \ref{eq:LLG} and
obtain

\begin{equation}
\begin{array}{c}
\vec{L}\rightarrow\vec{L}\left(\vec{b},\vec{m}\right)=-\gamma\vec{M}\left(t\right)\times\vec{B}(t)\\
-\frac{\alpha}{M}\vec{M}\left(t\right)\times\dot{\vec{M}}\left(t\right)-\dot{M}\left(t\right)\overset{!}{=}0\,\forall t
\end{array}\label{eq:LLG-Time}
\end{equation}
 which defines the hyper-plane in which all dynamic motion of the
magnetization takes place. Since $\vec{m}$ and $\vec{b}$ are small,
as defined in \ref{eq:Hf-Tensor-Def} we can now approximate $\vec{L}\left(\vec{b},\vec{m}\right)$
by using a Taylor-expansion around $\vec{L}\left(\vec{b}=\vec{0},\vec{m}=\vec{0}\right)$
to obtain
\begin{equation}
\vec{L}\left(\vec{b},\vec{m}\right)\approx\underset{\vec{0}}{\underbrace{\vec{L}\left(\vec{0},\vec{0}\right)}}+\underline{\underline{J}}_{\vec{b},\vec{m}}\cdot\left(b_{x},b_{y},b_{z},m_{x},m_{y},m_{z}\right)^{\top}
\end{equation}
This leads to the system of equations \ref{eq:LGSchi}, 
\begin{equation}
\vec{0}\overset{!}{=}\underline{\underline{J}}_{\vec{b},\vec{m}}\cdot\left(b_{x},b_{y},b_{z},m_{x},m_{y},m_{z}\right)^{\top}\label{eq:LGSchi}
\end{equation}
where $\underline{\underline{J}}_{\vec{b},\vec{m}}=\underline{\underline{J}}_{\left(b_{x},b_{y},b_{z},m_{x},m_{y},m_{z}\right)^{T}}\left(L\left(\vec{b},\vec{m}\right)\right)$
is the Jacobian matrix of $\vec{L}$ in $\vec{b}$ and $\vec{m}$.
Eq. \ref{eq:LGSchi} can then be further decomposed into 

\begin{equation}
\begin{aligned}\vec{0}= & \underline{\underline{J}}_{\vec{b}}\cdot\vec{b}+\underline{\underline{J}}_{\vec{m}}\cdot\vec{m}\\
\vec{m}= & -\left(\left(\underline{\underline{J}}_{\vec{m}}\right)^{-1}\cdot\underline{\underline{J}}_{\vec{b}}\right)\cdot\vec{b}
\end{aligned}
\end{equation}
where $\underline{\underline{J}}_{\vec{m}}$ and $\underline{\underline{J}}_{\vec{b}}$
are the Jacobian matrices in $\vec{m}$ and $\vec{b}$ respectively.
By comparison to eq. \ref{eq:Hf-Tensor-Def} we find 
\begin{equation}
\underline{\underline{\chi}}_{\mathrm{hf}}=-\left(\left(\underline{\underline{J}}_{\vec{m}}\right)^{-1}\cdot\underline{\underline{J}}_{\vec{b}}\right)\label{eq:HF-Tensor-Sol}
\end{equation}
which we refer to as the complete analytic solution of the ferromagnetic
high-frequency susceptibility. Note that this approach is independent
of the form of the free energy functional. Since we obtain the full
tensor without assumptions regarding its entries, we have to project
it on the unit vectors $\vec{u}_{b}$ and $\vec{u}_{m}$ that represent
an excitation-measurement-pair of observables to obtain a representative spectrum.
In a typical numerical evaluation one would assume $\vec{u}_{b}$ to
be parallel to the unit-vector in $\phi$ direction of the applied
field $\vec{B}$ and $\vec{u}_{m}$ to be parallel to the unit-vector
in $\phi$ direction of the magnetization vector in spherical coordinates.
Nonparallel unit-vectors $\vec{u}_{b}$ and $\vec{u}_{m}$ can be
used to account for nonuniform microwave fields. The angle between $\vec{u}_{b}$
and $\vec{u}_{m}$ represents an effective phase shift $\Delta$ between
the excitation and the response. This is illustrated in the inset
in fig. \ref{fig:Amplitude-of-the}. Such a phase shift can be created
for instance by having the sample covered by a conductive layer
in which the microwave creates an eddy current that in turn creates
a phase shifted microwave signal that superimposes with the original
one as described in \cite{Eddycurrent}. The approach presented here
was used in \cite{salikhov2015enhanced} to calculate asymmetric line
shapes. 

\subsection{Extension to reciprocal-($\vec{k}$-)space and description of the magnon dispersion}

The model presented above can be extended to reciprocal space in order
to obtain the magnon dispersion. Accordingly, the spatial contributions
to the energy landscape are included in the energy density
formulation. Also the Ansatz has to be changed such that the dynamic
magnetization has a spatial dependence. We imagine that the spatial
distribution of the magnetization can be described as a constant part
and a dynamic part where the dynamic part is a Fourier series. In
contrast to the description by Suhl, where this Ansatz appears \cite{SUHL1957209}
we consider the amplitude for every $\vec{k}$ to be small, such that
we can perturb the system with a single $\vec{k}$ at a time, yielding
an Ansatz of the form
\begin{equation}
\vec{M}\left(t,x\right)\coloneqq\vec{M}\left(M,\theta_{M},\varphi_{M}\right)+\vec{m_{k}}\exp\left(\imath\omega t-\vec{k}\cdot\vec{x}\right)
\end{equation}
where $\vec{k}$ is the reciprocal vector for which the susceptibility
is being calculated and $\vec{x}$ is the spatial coordinate at which
the wave is observed.

For example we can consider exchange energy contribution in a continuum
model 
\begin{equation}
F_{\mathrm{ex}}=d^{2}\frac{B_{ex}}{\left\Vert \vec{M}\right\Vert }\left(\vec{M}\left(t,x\right)\cdot\triangle\vec{M}\left(t,x\right)\right)
\end{equation}
and a $\vec{k}$ dependent dipolar coupling
to include dynamic aspects of dipolar interactions
\begin{equation}
F_{\mathrm{Demag}}=\frac{1}{2}\mu_{0}\left\Vert \vec{M}\right\Vert \frac{\vec{m}_{k}\cdot\vec{k}}{\left\Vert \vec{m}\right\Vert ^{2}\left\Vert \vec{k}\right\Vert ^{2}}\vec{k}\cdot\vec{M}\left(t,x\right)
\end{equation}
where $d$ is the distance between two neighboring spins, $B_{ex}$
is the exchange field they exert on each other and $\triangle$ is
the Laplace operator in real space.
\begin{figure}
\noindent \begin{centering}
\includegraphics[width=0.5\columnwidth]{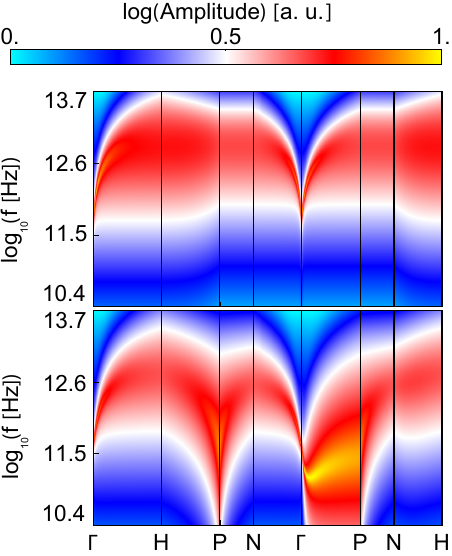}
\par\end{centering}
\caption{The magnon dispersion calculated for a bcc structure including a cubic
anisotropy exchange and dipolar coupling (top) and for a similar
system with an additional strong chiral \cite{Dzyaloshinskii1957,Moriya1960} coupling (bottom). \label{fig:The-magnon-dispersion}}

\end{figure}
 Adding this contribution to the Helmholtz energy density we can proceed
as before and calculate the susceptibility for every $\vec{k}$ in
the Brillouin zone as shown exemplary in fig. \ref{fig:The-magnon-dispersion}. The results agree with the literature (see for instance \cite{Keffer1966,Kittel1963,Moon2013,PhysRevB.91.214409,RezendeYIGField}).
For other spatial contributions such as anisotropic exchange and chiral
coupling , the model can be applied in the same way.

\subsection{The equilibrium position of the magnetization\label{subsec:The-equilibrium-position}}

In order to use the result in eq. \ref{eq:HF-Tensor-Sol} to obtain
the susceptibility it is necessary for $\vec{M}\left(\theta_{M},\phi_{M}\right)$
to locally minimize the free energy density. The orientation of the
magnetization vector has to be determined form the shape of the free
energy landscape including an applied magnetic field. In the following
we present our recursive method to efficiently find these minima.
In terms of infinitesimals this method can be viewed 
as a trajectory depended analytic solution. Due to its infinite recursion
along a chosen trajectory however it resembles a second order newton
algorithm, which is a numerical tool, and we therefore tend to call
it a semi-analytic trajectory dependent solution of the equilibrium
states of the magnetization.

\begin{figure}
\noindent \begin{centering}
\includegraphics[width=0.5\columnwidth]{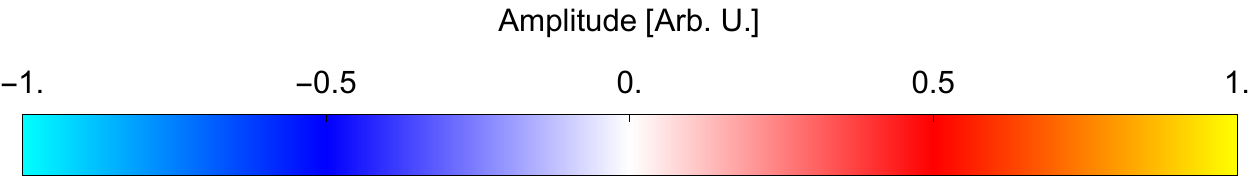}
\par\end{centering}
\noindent \begin{centering}
\includegraphics[width=0.25\columnwidth]{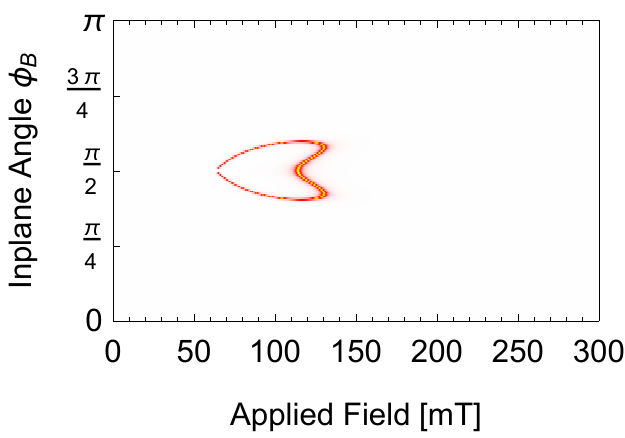}\includegraphics[width=0.25\columnwidth]{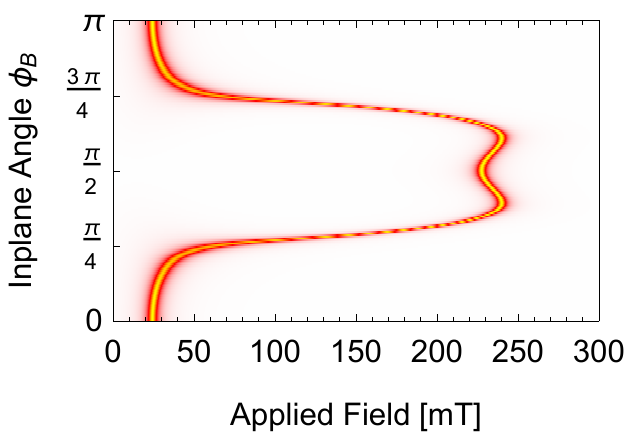}
\par\end{centering}
\noindent \begin{centering}
\includegraphics[width=0.5\columnwidth]{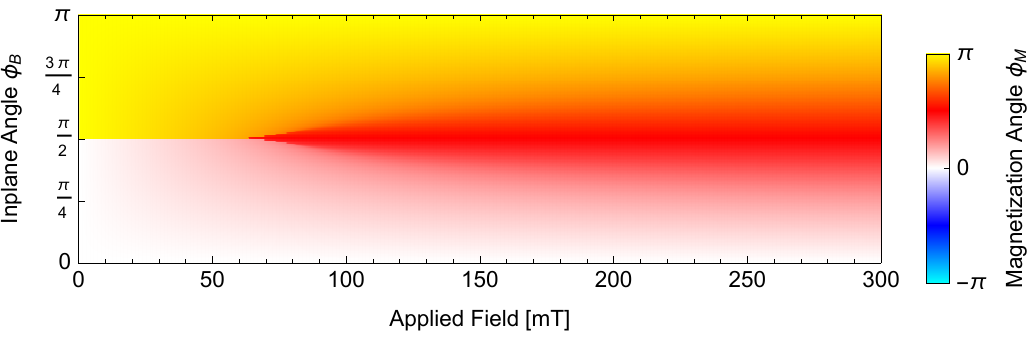}
\par\end{centering}
\caption{Calculated spectra at typical X-Band frequencies: 10 GHz (top left),
and 18GHz (top right) and the corresponding solutions for the magnetization
angles (bottom). The model that has been used for the free energy
here is a cubic anisotropy $K_{4}=\unit[4.8\cdot10^{4}]{J/m^{3}}$
where the 110-direction is perpendicular to the azimuthal plane ($\theta=\frac{\pi}{2}$)
together with an in-plane uni-axial anisotropy $K_{u}=\unit[-7.5\cdot10^{4}]{J/m^{3}}$
and a 2-fold out of plane anisotropy $K_{2}=\unit[0.3\cdot10^{4}]{J/m^{3}}$
according to \cite{0034-4885-61-7-001} with the demagnetizing tensor
of a thin film. The g-factor was set to $2.09$ and the damping constant
of $\alpha=0.004$ was used.\label{fig:Calculated-spectra-at}}

\end{figure}

For certain paths in the applied field space, where the trajectory passes sufficiently far beyond a hard direction, the equilibrium angles
are discontinuous if the Zeeman energy does not overcome the anisotropic
contributions of this hard direction. This can lead to a hysteretic
behavior of the magnetization depending on the trajectory of $\vec{B}\left(\tau\right):=\vec{B}\left(B\left(\tau\right),\theta_{B}\left(\tau\right),\varphi_{B}\left(\tau\right)\right)$.
To account for this behavior a solution representing the equilibrium
angles must depend on the trajectory $\vec{B}\left(\tau\right)$ and
not only on a momentary configuration of $\vec{B}$. Without loss
of generality we will only consider the equilibrium angles $\left\{ \theta_{M},\varphi_{M}\right\} $
of the magnetization in spherical coordinates to minimize the free
energy, since in many applications the norm of the magnetization may
be considered constant. Once a minimizer $\vec{\Omega}\left(\vec{B}\left(0\right)\right)=\left\{ \theta_{M},\varphi_{M}\right\} _{0}$
of the free energy $F\left(\vec{B},\vec{M}\right)$ is known for a
certain starting configuration $\vec{B}\left(0\right)$, a small change
in $\vec{B}\rightarrow\vec{B}\left(0+\delta\right)$ that yields a
small change in the position of the minimum of $F\left(\vec{B},\vec{M}\right)$
can be accounted for by calculating a series expansion of $F\left(\vec{B}\left(0+\delta\right),\vec{M}\right)$
at the position $\left\{ \theta_{M},\varphi_{M}\right\} _{0}$ to
the second order. 
\begin{figure}
\noindent \begin{centering}
\includegraphics[width=0.5\columnwidth]{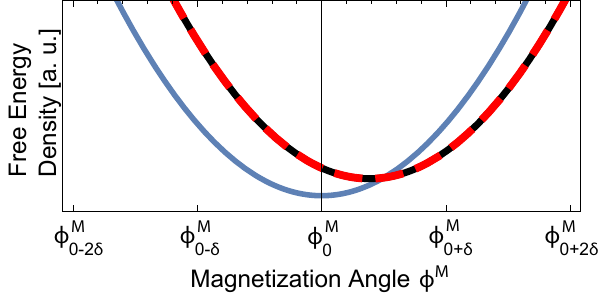}
\par\end{centering}
\caption{The environment of a minimum of the energy Landscape as a function
of the Magnetization angle (blue) and the same energy landscape after
changing the applied field angle $\phi_{B}$ by a small quantity $\delta$
(black) together with a Taylor expansion of the changed energy landscape
around the the position $\phi_{0}^{M}$ (red dashed). Note that the series 
expansion fits perfectly to the black curve around its minimum.\label{fig:The-ennviroment-of}}

\end{figure}
The position of the minimum of this parabola will be close to the
minimum $\left\{ \theta_{M},\varphi_{M}\right\} _{0+\delta}$ of $F\left(\vec{B}\left(0+\delta\right),\vec{M}\right)$.
In fact as $\delta$ decreases the solution obtained this way will
get closer to the exact minimum. This procedure is illustrated in
fig. \ref{fig:The-ennviroment-of}, where the free energy was defined to be $F=\sin^{2}\left(2\phi_{M}\right)-5\cos\left(\phi_{B}-\phi_{M}\right)$.
Since the function obtained from the series expansion is of quadratic
order it can always be written in a form such that the vertex can
be directly extracted from the function. Therefore a recursive function
of the form \ref{eq:Rec-Form}
\begin{align}
\vec{\Omega}\left(\vec{B}\left(0-\delta\right)\right)=\vec{\Omega}\left(\vec{B}\left(0\right)\right)-\nonumber \\
\left.\underline{\underline{H}}_{F}^{-1}\right|_{\vec{\Omega}\left(\vec{B}\left(0-\delta\right)\right)}\cdot\left.\vec{\nabla}F\right|_{\vec{\Omega}\left(\vec{B}\left(0-\delta\right)\right)}\nonumber \\
\vec{\Omega}\left(\vec{B}\left(0-2\delta\right)\right)=\vec{\Omega}\left(\vec{B}\left(0-\delta\right)\right)-\label{eq:Rec-Form}\\
\left.\underline{\underline{H}}_{F}^{-1}\right|_{\vec{\Omega}\left(\vec{B}\left(0-2\delta\right)\right)}\cdot\left.\vec{\nabla}F\right|_{\vec{\Omega}\left(\vec{B}\left(0-2\delta\right)\right)}\nonumber \\
...\nonumber 
\end{align}
 can be derived to describe the position of a minimum for certain
trajectories $\vec{B}\left(\tau\right)$, where $\underline{\underline{H}}_{F}$
is the Hessian Matrix of the free energy density that described the
curvature and $\vec{\nabla}F$ the gradient that describes the slope
of the free energy. Conceptually this can be considered a second order
Newton algorithm with the exception that it starts from a known position
making the number of iterations required tend towards $1$ as $\delta$
gets small. To determine a minimizer that can be used as a starting
point in eq. \ref{eq:Rec-Form} the easiest approach in a numerical
calculation is to start at a field value sufficiently higher than
the field at which the Zeeman energy fully overcomes the anisotropy
energy \textendash{} in the sense that there is only one minimum and
one maximum left in the energy landscape \textendash{} and to assume
that the magnetization is parallel to the applied field in this configuration.
This approach was implemented and found to be very accurate in \cite{salikhov2015enhanced}
for fitting FMR spectra recorded at different microwave frequencies.
Figure \ref{fig:Calculated-spectra-at} shows some calculated spectra
using the solution presented above, with the corresponding equilibrium
angles calculated with this trajectory dependent algorithm. The overall
calculation time was about five minutes for $540180$ data points.

Equation \ref{eq:HF-Tensor-Sol} in combination with eq. \ref{eq:Rec-Form}
describe a very fast algorithm to calculate the complete susceptibility
for any given free energy density and any measurement trajectory.
This algorithm however will not always align the magnetization in
the absolute minimum of the free energy, in fact it will fall into
meta-stable states if for instance a fourfold crystalline anisotropy
is considered and the applied field is swept along the field angle
rather than the field amplitude, predicting the occurrence of ferromagnetic
resonance in meta-stable states. 

\section{Summary}

We have devised a versatile analytic model, capable of accurately
describing FMR experiments as well as modeling the full magnon dispersion.
The model is simple in that it requires only derivatives. Condensed
into a single operator $\underline{\underline{\chi}}_{\mathrm{hf}}$,
it is compact and thus easy to use in analytic and numeric applications.
The formulation through an energy density allows for easy modification
of the model to adapt different types of interactions, such as dipole-dipole-interaction,
spin-spin-interactions like the Dzyaloshinski\v{\i}-Moriya interaction
and spin-orbit interactions. It can also be applied directly to spatial
dependent spin configurations obtained from micromagnetic simulations
to retrieve information about the magnetodynamic properties of spin
textures. The model is not restricted to evaluating the magnon dispersion
as a function $\omega\left(k\right)$ but instead yields the magnonic
response amplitude $\chi\left(\omega,k\right)$ as a Green's function.
In addition to this, the algorithm described in sec. \ref{subsec:The-equilibrium-position}
makes it possible to apply the model on orientations of the magnetization
which are non collinear with the symmetry directions of the system
or the applied magnetic field. This can be used to calculate angular
dependent spectra, as well as identify meta-stable states and describe
their magnetodynamic behavior.

\bibliography{literature}

\end{document}